\documentclass[twoside]{book}% Specify document type
\usepackage{graphicx}       % Load macros for PostScript figures
 \makeatletter
\ifx\UNDEF\mail\def\mail{ }\else\fi
\ifx\UNDEF\prange\def\prange{0 0}\else\fi

\gdef\@empty{}
\def\Mail#1 #2 {\gdef\thecontact{#1}\gdef\theaddr{#2}}
\def\Range#1 #2 {\gdef\thefirstpage{#1}\gdef\thelastpage{#2}}
{\let\'\mail \expandafter\Mail\' }	% do not remove space between ' and }
{\let\'\prange \expandafter\Range\' }	% do not remove space between ' and }
 \gdef\@shtitle{\relax}
 \long\def\shtitle#1{\gdef\@shtitle{#1}}
 \long\def\author#1{\gdef\@author{#1}}
 \def\affil#1{\par\noindent{\rm#1\par}}
 \gdef\@abstract{}
 \long\def\abstract#1{\gdef\@abstract{#1}}

 \renewcommand{\@evenhead}{\thepage\qquad\qquad\@shtitle\hfil}
 \renewcommand{\@oddhead}{\hfil\@shtitle\qquad\qquad\thepage}

 \def\maketitle{\thispagestyle{empty}\chapter{\@title}}
 \renewcommand\chapter{\if@openright\cleardoublepage\else\clearpage\fi
                    \thispagestyle{empty}%
                    \global\@topnum\z@
                    \@afterindentfalse
                    \secdef\@chapter\@schapter}
 \def\@makechapterhead#1{%
  \vspace*{50\p@}%
  {\parindent \z@ \raggedleft \normalfont
    \ifnum \c@secnumdepth >\m@ne
      \if@mainmatter
        \huge \@chapapp{} \thechapter
        \par\nobreak
        \vskip 20\p@
      \fi
    \fi
    \interlinepenalty\@M
    \Huge \bfseries #1\par\nobreak
    \vskip.25in
    \large\bfseries\@author\par\nobreak
    \vskip 40\p@}
    \ifx\@abstract\@empty\else{\small\@abstract\par\vskip20\p@}\fi
  }

\DeclareRobustCommand\em
        {\@nomath\em \ifdim \fontdimen\@ne\font >\z@
                       \upshape \else \slshape \fi}

\def\@begintheorem#1#2{\sl \trivlist \item[\hskip \labelsep{\bf #1\ #2}]}
\def\@opargbegintheorem#1#2#3{\sl \trivlist
     \item[\hskip \labelsep{\bf #1\ #2\ (#3)}]}

  \def\@arabic#1{\number #1} % my redefinition

\long\def\@makecaption#1#2{
	\vskip\abovecaptionskip
	\sbox\@tempboxa{{\small {\bf #1}: #2}}%
	\ifdim\wd\@tempboxa>\hsize
	    {\small {\bf #1}: #2\par}
	\else
	   \global\@minipagefalse
	   \hbox to\hsize{\hfil\box\@tempboxa\hfil}
	\fi
	\vskip \belowcaptionskip}

\def\figstrut#1{\hbox to\linewidth{\vrule height#1\hfill}}

\renewenvironment{thebibliography}[1]
     {\section*{\bibname
        \@mkboth{\MakeUppercase\bibname}{\MakeUppercase\bibname}}%
      \list{\@biblabel{\@arabic\c@enumiv}}%
           {\settowidth\labelwidth{\@biblabel{#1}}%
            \leftmargin\labelwidth
            \advance\leftmargin\labelsep
            \@openbib@code
            \usecounter{enumiv}%
            \let\p@enumiv\@empty
            \renewcommand\theenumiv{\@arabic\c@enumiv}}%
      \sloppy
      \clubpenalty4000
      \@clubpenalty \clubpenalty
      \widowpenalty4000%
      \sfcode`\.\@m}
     {\def\@noitemerr
       {\@latex@warning{Empty `thebibliography' environment}}%
      \endlist}
\makeatother

\shtitle{Multidimensional Network Monitoring}  % short title line
 \title{Multidimensional Network Monitoring\\ for Intrusion Detection}
  \author{Vladimir Gudkov and Joseph E. Johnson\affil{Department of Physics and Astronomy \\
 University of South Carolina \\
 Columbia, SC 29208\\gudkov@sc.edu;  jjohnson@sc.edu}}
 \abstract{An approach for real-time network monitoring in terms of numerical
time-dependant functions of protocol parameters is suggested.
Applying complex systems theory for information f{l}ow analysis of
networks, the information traffic is described as a
trajectory in multi-dimensional parameter-time space with about
10-12 dimensions. The network traffic description is
synthesized by applying methods of theoretical physics and complex
systems theory, to provide a robust approach for network
monitoring that detects known intrusions, and supports developing
real systems for detection of unknown intrusions. The methods of
data analysis and pattern recognition presented are the basis of a
technology study for an automatic intrusion detection system that
detects the attack in the reconnaissance stage.}

\begin{document}           % Matched by \end{document}
\maketitle

\section{Introduction}

Understanding the behavior of an information network and
describing its main features are very important for information
exchange protection on computerized information systems. Existing
approaches for the study of network  attack tolerance usually
include the study of the dependance of network stability on
network complexity and topology (see, for example
\cite{cnet1,cnet2} and references therein); signature-based
analysis technique; and statistical analysis and modelling of
network traffic (see, for example \cite{sig1,st1,st2,st3}).
Recently, methods to study spatial traffic flows\cite{spflow} and
correlation functions of irregular sequences of numbers occurring
in the operation of computer networks \cite{timetr} have been
proposed.

Herein we discuss properties related to information exchange on
the network rather than network structure and topology. Using
general properties of information flow on a network we suggest a
new approach for network monitoring and intrusion detection, an
approach based on complete network monitoring. For detailed
analysis of information exchange on a network we apply methods
used in physics to analyze complex systems. These methods are
rather powerful for general analysis and provide a guideline by
which to apply the result for practical purposes such as real time
network monitoring, and possibly, solutions for real-time
intrusion detection\cite{indet}.

\section{Description of Information Flow}

A careful analysis of information exchange on networks leads to
the appropriate method to describe information flow in terms of
numerical functions. It gives us a mathematical description of the
information exchange processes, the basis for network simulations
and analysis.

To describe the information flow on a network, we work on the
level of packet exchange between computers.  The structure of the
packets and their sizes vary  and depend on the process. In
general, each packet consists of a header and attached
(encapsulated) data. Since the data part does not affect packet
propagation through the network, we consider only information
included in headers. We recall that the header consists of
encapsulated protocols related to different layers of
communications, from a link layer to an  application layer. The
information contained in the headers controls all network traffic.
To extract this information one uses tcpdump utilities developed
with the standard of LBNL's Network Research Group \cite{tcpd}.
 This information is used to analyze network traffic to
find a signature of abnormal network behavior and to detect
possible intrusions.

The important difference of the proposed approach from
traditionally used methods is the presentation of information
contained in headers in terms of well-defined numerical functions.
To do that we have developed software to read binary tcpdump files
and to represent all protocol parameters as corresponding
time-dependent functions. This gives us the opportunity to analyze
complete information (or a chosen fraction of complete information
that combines some parameters) for a given time and time window.
The ability to vary the time window for the analysis is important
since it makes possible extracting different scales in the time
dependance of the system. Since different time scales have
different sensitivities for particular modes of system behavior,
the time scales could be sensitive to different methods of
intrusion.

As  was done in reference paper\cite{gj1}, we divide the protocol
parameters for host-to-host communication  into two separate
groups with respect to the preserving or changing their values
during packet propagation through the network (internet). We refer
to these two groups of parameters as ``dynamic'' and ``static''.
The dynamic parameters may be changed during packet propagation.
For example, the ``physical'' address of a computer, which is  the
MAC parameter of the Ethernet protocol, is a dynamic parameter
because it can be changed if the packet has been re-directed by a
router. On the other hand, the source IP address is an example of
a static parameter because its value does not change during packet
propagation. To describe the information flow, we use only static
parameters since they may carry intrinsic properties of the
information flow and neglect the network (internet) structure. (It
should be noted that the dynamic parameters may be important for
study of network structure properties. Dynamic parameters will be
 considered separately.)

Using packets as a fundamental object for information exchange on
a network and being able to describe packets in terms of functions
of time for static parameters to analyze network traffic, we can
apply methods developed in physics and applied mathematics to
study dynamic complex systems. We present some results obtained in
references \cite{gj1,gj2} to demonstrate the power of these
methods and to recall important results for network monitoring
applications.

It was shown \cite{gj1} that to describe information flow on a
network one can use a small number (10 - 12) of parameters. In
other words, the dimension of the information flow space is less
than or equal to 12 and the properties of information flow are
practically independent of network structure, size and topology.
To estimate the dimension of the information flow on the network
one can apply the algorithm for analysis of observed chaotic data
in physical systems, the algorithm suggested in paper \cite{abar1}
(see also ref. \cite{chaos}and references therein). The main idea
 relates to the fact that any dynamic system with dimensionality
of $N$ can be described by a set of $N$ differential equations of
the second order in configuration space or by a set of $2N$
differential equations of first order in phase space.

Assuming that the information flow can be described in terms of
ordinary differential equations (or by discrete-time evolution
rules), for some unknown functions  in a (parametric) phase space,
one can analyze a time dependance of a given scalar parameter
$s(t)$ that is related to the system dynamics. Then one can build
$d$-dimensional vectors from the variable $s$ as
\begin{equation}
y^d(n)=[s(n),s(n+T),s(n+2T),\ldots , s(n+T(d-1))] \label{yvec}
\end{equation}
 at equal-distant time intervals $T$:
 $s(t) \rightarrow s(T\cdot n) \equiv s(n)$, where $n$ is
an integer number to  numerate $s$ values at different times. Now,
one can calculate a number of nearest neighbors in the vicinity of
each point in the vector space and plot the dependance of the
number of false nearest neighbors (FNN) as a function of time. The
FNN for the $d$-dimensional space are neighbors that move far away
when we increase dimension from $d$ to $d+1$ (see, for details
ref.\cite{gj1}).

\begin{figure}
\includegraphics{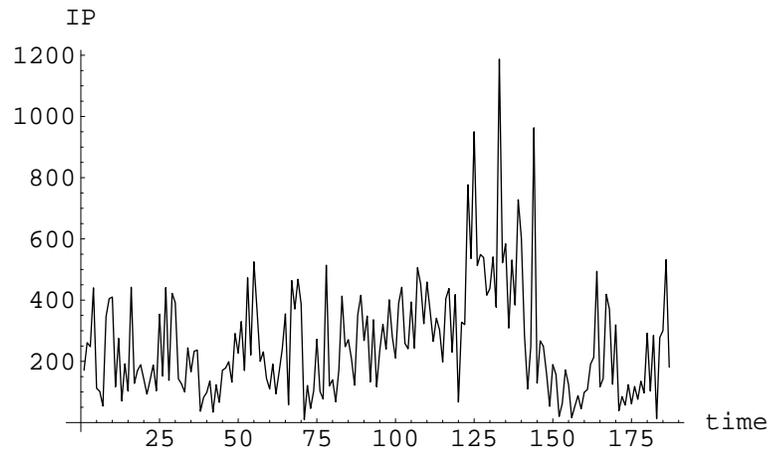}
\caption{Protocol type ID in the IP protocol as a function of time
(in $\tau = 5 sec$ units).} \label{fig:ip}
\end{figure}

\begin{figure}
\includegraphics{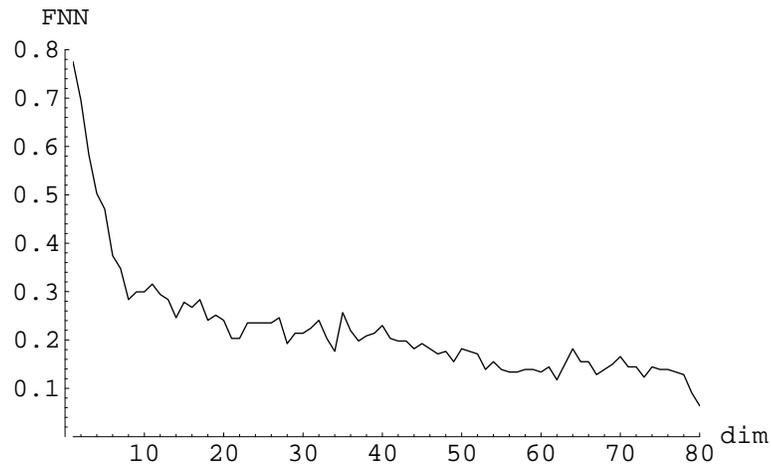}
\caption{Relative number of false nearest neighbors as a function
of dimension of unfolded space.} \label{fig:fnn}
\end{figure}

The typical behavior of a scalar parameter and corresponding FNN
plot are shown in Figs. (\ref{fig:ip}) and (\ref{fig:fnn}).
From the last plot one can see that the number of FNN rapidly
decreases  up to about 10 or 12 dimensions. After that it shows a
slow dependency on the dimension, if at all. Fig. (\ref{fig:fnn})
shows that by increasing the dimension $d$ step-by-step, the
number of FNN, which occur due to projection of far away parts of
the trajectory in higher dimensional space is decreases  to a
level restricted by  system noise that has infinite dimension.
Therefore, for a complete description of the information flow one
needs not more than 12 independent parameters. The dynamics of
information flow can be described as a trajectory in a phase space
with the dimension of about 10 - 12. Since this dimension does not
depend on the network topology, its size, and the operating
systems involved in the network, this is a universal
characteristic and may be applied for any network.

\begin{figure}
\includegraphics{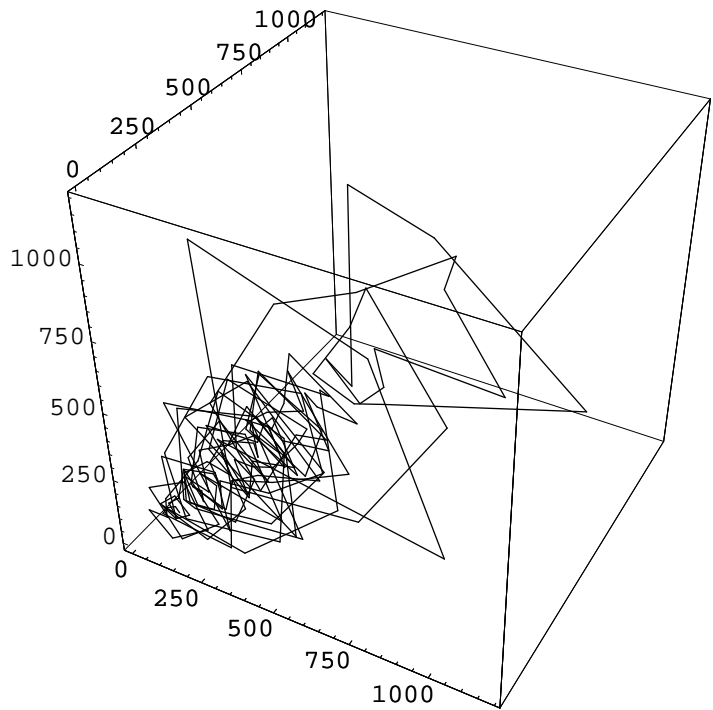}
\caption{The projection of the trajectory of the information flow
3-dimensional phase space.} \label{fig:atr}
\end{figure}

However, we cannot identify exactly these independent parameters.
Due to the complexity of the system it is natural that these
unknown parameters which are real dynamic degrees of freedom of
the system would have a complicated relationship with the
parameters contained in the network protocols.  Fortunately, the
suggested technique provides very powerful methods to extract
general information about the behavior of dynamic complex systems.
For example, the obtained time dependence of only one parameter,
the protocol ID shown on Fig.(\ref{fig:ip}), is enough to
reconstruct the trajectory of the information flow in its phase
space. The reconstructed projection of the trajectory on
3-dimensional space is shown on Fig. (\ref{fig:atr}). Therefore,
one can see that the complete description of the network
information traffic in terms of a small number of parameters is
possible. The important point is that this trajectory  (usually
called as an ``attractor'') is
 well-localized.  Therefore, it can be used for detailed analysis
and pattern recognition techniques. It should be noted that the
attractor presented here is obtained from one parameter
measurement only, for that being illustrative purposes. For real
analysis we use multi-dimensional high accuracy reconstruction.

\section{Real Time Network Monitoring and Detection of Known Intrusions}

The proposed approach for network traffic description provides the
possibility of real-time network monitoring and detection of all
known network attacks. This is because one collects from tcpdump
binary output the complete information about network traffic at
any given point in the network. All header parameters are
converted into time dependant numerical functions. Therefore, each
packet for host-to-host exchange corresponds to a point in the
multidimensional parametric phase space. The set of these points
(the trajectory) completely describes information transfer on the
network. It is clear that this representation provides not only
the total description of the network traffic at the given point
but also a powerful tool for analysis in real time. Let us
consider some possible scenarios for the analysis.

Suppose we are looking for known network intrusions. The signature
of an intrusion is a special set of relationships among the header
parameters. For example \cite{indet}, the signature for the
attempt to identify live hosts by those responding to the ACK scan
includes a source address, an ACK and SYN flags from TCP protocol,
a target address of the internal network, sequence numbers, and
source and destination port numbers. The lone ACK flag set with
identical source and destination ports is the signature for the
ACK scan.  This is because the lone ACK flag set should be found
only as the final transmission of the three-way handshake, an
acknowledgement of receiving data, or data that is transmitted
where the entire sending buffer has not been emptied. From this
example one can see that the intrusion signature could be easily
formulated in terms of logic rules and corresponding equations.
Then, collecting the header parameters (this is the initial phase
of network monitoring) and testing sets of them against the
signatures (functions in terms of the subset of the parameters)
one can filter out all known intrusions. Since we can collect any
set of the parameters and easily add any signature function, it
provides the way for a continuous upgrading of the intrusion
detection system (IDS) built on these principles.  In other words,
such an IDS is universal and can be used to detect all possible
network intrusions by adding new filter functions or macros in the
existing testing program. It is very flexible and easily
upgradable.  The flexibility is important  and can be achieved
even in existing ``traditional'' IDS's. What is out of scope of
traditional approaches is the mathematically optimized
minimization of possible false alarms and controlled sensitivity
to intrusion signals. These properties are an intrinsic feature of
our approach.

The important feature of the approach is the presentation of the
parameters in terms of time dependant functions. This gives the
opportunity to decrease  as best as possible for the particular
network the false alarm probability of the IDS. This can be done
using
 sophisticated methods already developed for noise reduction in time series.
Moreover, representation of the protocol parameters as numerical
functions provides the opportunity for detailed mathematical
analysis and for the optimization of the signal-to-noise ratio
using not only time series techniques but also numerical methods
for analysis of multi-dimensional functions. The combination of
these methods provides the best possible way, in terms of accuracy
of the algorithms and reliability of the obtained information, to
detect of known intrusions in real time.

Also, the description of the information flow in terms of
numerical functions gives the opportunity to monitor network
traffic for different time windows without missing information and
without overflowing storage facilities. One can suggest
 ways to do it. One example is the use of a parallel
computer environment (such as low cost powerful Linux clusters)
for the simultaneous analysis of the decoded binary tcpdump
output. In this case the numerical functions of the header
parameters being sent to different nodes of the cluster will be
analyzed by each node using similar algorithms but different
scales for time averaging of signals (or functions). Thus, each
node has a separate time window and, therefore, is sensitive to
 network behavior in the particular range of time. For example, choosing
time averaging scales for the nodes from microseconds to weeks,
 one can trace and analyze network traffic
independently and simultaneously in all these time windows. It is
worthwhile to remember that the optimal signal-to-noise ratio is
achieved for each time window independently thereby providing the
best possible level of information traffic analysis for the whole
network. There are  three obvious advantages for this approach.
The first is the possibility to detect intrusions developed on
different time scales simultaneously and in real time. The second
is the automatic decreasing of noise from short time fluctuations
for long time windows due to time averaging. This provides
detailed information analysis in each time window without loss of
information. At the same time, it discards all noise related
information, drastically reducing the amount of information at the
storage facilities. The third advantage is the possibility to use
(in real time) the output from short time scale analyzed data as
additional information for long time scale analysis.

\begin{figure}
\includegraphics{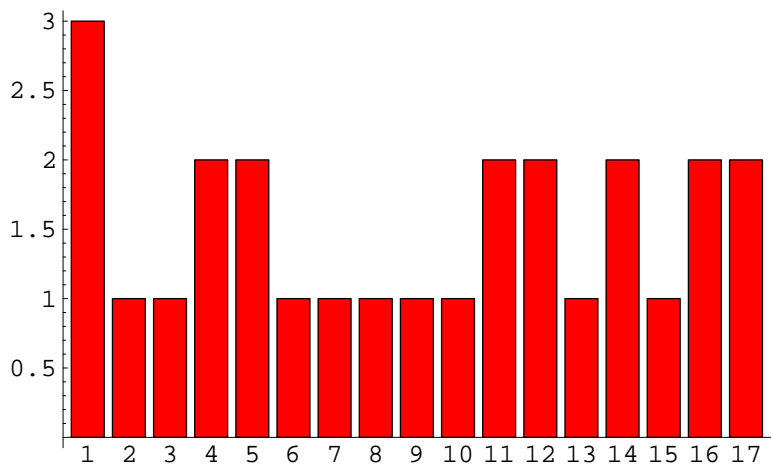}
\caption{Frequencies of the parameters used in signatures of
intrusions. For numbering rules see Table 1.} \label{fig:fr}
\end{figure}

To give an idea of how many parameters are used to describe
signatures of currently known intrusions we use the result of  the
comprehensive (but probably not complete) analysis\cite{gj2}  of
known attacks, i.e., smurf, fraggle, pingpong, ping of death, IP
Fragment overlap, BrKill , land attack , SYN flood attack, TCP
session hijacking, out of band bug, IP unaligned timestamp, bonk,
OOB data barf, and vulnerability scans (FIN and SYN \& FIN
scanning). The frequencies of the parameters involved in
signatures for these intrusions are shown on Fig.(\ref{fig:fr}).
 The numeration of the parameters is explained in Table 1. One
can see that the number of parameters  used for signatures of
intrusions is rather small . This fact further
 simplifies the procedure of the analysis. \\

\begin{table}
\caption{The parameters involved in intrusion signatures as shown
on Fig.(\ref{fig:fr}).}
 \begin{tabular}{|c|c|c|c|c}
 \hline
 Number & Protocol & Parameter & Frequency \\
 \hline

1 &   IP & Destination IP Address  & 3 \\
2 &  IP & Source IP Address  &  1 \\
3 &  IP & Length  & 1 \\
4 & IP & More Fragment Flag &  2 \\
5 & IP & Don't Fragment Flag & 2\\
6 &  IP & Options & 1 \\
7 & TCP & Source Port & 1 \\
8 & TCP & Destination Port &  1 \\
9 & TCP & Urgent Flag & 1 \\
10 & TCP & RST Flag   & 1 \\
11 & TCP & ACK Flag   & 2 \\
12 & TCP & SYN Flag   & 2 \\
13 & TCP & FIN Flag   & 1 \\
14 & UDP & Destination Port  &  2 \\
15 & UDP & Source Port & 1 \\
16 & ICMP & Type & 2 \\
17 & ICMP & Code &  2 \\
\hline
 \end{tabular}
 \end{table}

\section{Detection of Unknown Intrusions}

The aforementioned approach could be considered a powerful and
promising method for network monitoring and detection of known
network intrusions.  However, the more important feature of the
approach is the ability to detect previously unknown attacks on a
network in a wide range of time scales. This ability is based on
the method of describing information exchange on a network in
terms of numerical functions of header parameters (or a trajectory
in multi-dimensional phase space) as well as using methods of
theoretical physics for the analysis of dynamics of complex
systems. These methods lead to a very useful
 result for the small dimensionality of the
information flow space. Since the number of parameters used in
packet header is large (on the order of hundreds), the practical
search for unknown (even very abnormal) signals would be a
difficult problem, if not impossible. Therefore, the small
dimension of the parametric space of the information flow is a
crucial point for the practical approach for the detection of
unknown intrusions.

To build a real time intrusion detection system that is capable of
detecting unknown attacks, we exploit the fact that we need to
analyze only a small number of parameters. Furthermore, as is
known from complex systems theory, the choice of the parameters is
not important unless they are sensitive to system behavior. The
last statement needs to be explained in more detail. Generally,
hundreds different parameters could be encapsulated in the packet
headers. The question is which parameters we need to choose for
the right description of the information flow. Following the
discussion in the previous section, one might surmise that we need
to make our choice from the known quoted 17 parameters. It may be
a good guess. However, the number 17 is  bigger than the dimension
of the phase space which we have in mind, and it could be that
hackers will invent new attacks with new signature parameters that
are not included in the set presented in the previous section. The
right answer to these remarks follows from complex systems theory.
For a complete system description one needs only the number of
parameters equal to the phase space dimension (more precisely, the
smallest integer number that is larger than fractal dimension of
the phase space). It could be a set of any parameters that are
sensitive to the system dynamics (and the 17 discussed parameters
could be good candidates). We do not know, and do not suppose to
know, the real set of parameters until the theory of network
information flow is developed or a reliable model for information
flow description is suggested. Nevertheless, a method developed to
study non-linear complex systems provides tools to extract the
essential information about the system from the analysis of even a
small partial set of the ``sensitive'' parameters. As an example,
one can refer to the Fig.(\ref{fig:atr}) which shows the
3-dimensional projection of the reconstructed trajectory from the
time dependent behavior of only one parameter (the protocol ID
shown on Fig.(\ref{fig:ip})). It means that the complete
description of the network information flow could be obtained even
from a small set of ``sensitive'' parameters.

One of the ways to implement this approach is to use the
multi-window method discussed in the previous section with the
proper data analysis for each time scale. This method of analysis
is not within the scope of the current paper and will be reported
elsewhere. We will review only the general idea and the problems
related to this analysis.  To detect unknown attacks (unusual
network behavior) we use a deviation of signals from the normal
regular network behavior.  For these purposes one can use a
pattern recognition technique to establish patterns for normal
behavior and  to measure a possible deviation from this normal
behavior. However, the pattern recognition problem is quite
difficult for this multidimensional analysis. According to our
knowledge, it is technically impossible to achieve reliable
efficiency in a pattern recognition for  space with  a rather
large dimension, such as  10. On the other hand, the more
parameters we analyze the better accuracy and reliability we can
 obtain. Therefore, we have to choose the optimal (compromise)
solution that uses pattern recognition techniques in information
flow subspaces with low dimensions. By applying appropriate
constraints on some header parameters one can choose these
subspaces as cross sections of the total phase space defined. In
this case, we will have a reasonable ratio of signal-to-noise and
will simplify the pattern recognition technique and improve its
reliability. For a pattern recognition we suggest using a 2-3
dimension wavelet analysis chosen on the basis of detailed study
of the information traffic on the set of networks. The wavelet
approach is promising because it reduces drastically and
simultaneously the computational time and memory requirements.
This is important for multidimensional analysis because it can be
used for an additional, effective noise reduction technique.

\section{Conclusions}

We suggest a new approach for multidimensional real time network
monitoring that is based on the application of complex systems
theory for information flow analysis of networks. Describing
network traffic in terms of numerical time dependant functions and
applying methods of theoretical physics for the study of  complex
systems provides a robust method for network monitoring to detect
known intrusions and is promising for development of real systems
to detect unknown intrusions.

To effectively apply innovative technology approaches
  against practical attacks it is necessary to detect and identify
  the attack in a reconnaissance stage.
 Based on new methods of data analysis and pattern
recognition, we are studying a technology to build an automatic
intrusion detection system. The system will be able to help
maintain a high level of confidence in the protection of networks.

We thank the staff of the Advanced Solutions Group for its
technical support.
 This work was supported by the DARPA Information Assurance and
 Survivability Program and is administered by the USAF Air Force
 Research Laboratory via grant F30602-99-2-0513, as modified.

\end{document}